%% file: gupta.tex
\documentclass{svmult}
\usepackage{epsfig,graphicx} 
\usepackage{amsmath}
\usepackage[bottom]{footmisc}
\usepackage{natbib,url}      
\bibpunct{(}{)}{;}{a}{}{,}   
\def\cite#1{\citealp{#1}}    

\begin{document}


\input{rr-latexdefs}  

\title*{On the statistical detection of propagating waves in polar coronal hole}


\author{G. R. Gupta\inst{1,2}
        \and
        E. O'Shea\inst{3}
        \and
        D. Banerjee\inst{1}
        \and
        M. Popescu\inst{3}
        \and
        J. G. Doyle\inst{3}}


\authorrunning{Gupta et al.}  

\institute{Indian Institute of Astrophysics, Bangalore, India
           \and 
           Indian Institute of Science, Bangalore, India
           \and
           Armagh Observatory, College Hill, Armagh, N. Ireland}
\maketitle

\setcounter{footnote}{0}  
\begin{abstract} 
Waves are important for the heating of the solar corona and the
acceleration of the solar wind. We have examined a long spectral time series of a
southern coronal hole observed on the 25th February 1997, with the SUMER
spectrometer on-board SoHO. The observations were obtained in a transition
region N~{\sc iv} 765 \AA\ line and in a low coronal Ne~{\sc viii} 770 \AA\ line. Our
observations indicate the presence of compressional waves with periods
of ~18 min. We also find significant power in shorter periods. Using Fourier techniques, we measured the phase delays between
intensity as well as velocity oscillations in the two chosen lines over a frequency domain. From
this we are able to measure the travel time of the propagating
oscillations and, hence, the propagation speeds of the waves producing the
oscillations. As the measured propagation speeds are subsonic, we conclude that the 
detected waves are slow magneto-acoustic in nature.

\end{abstract}

\section{Introduction}      \label{gupta-sec:introduction}

Coronal holes are regions of cool and low density plasma that are
`dark' at coronal temperatures. The predominantly unipolar
magnetic field from coronal hole regions is thought to give rise
to the fast solar wind.
A number of studies \eg\ \cite{1997ApJ...491L.111O, 2000ApJ...529..592O, 2001A&A...380L..39B, 2005A&A...442.1087P} 
have measured oscillations in coronal holes in the polar off-limb regions of
the Sun. All of these studies point to the presence of compressional waves, thought to
be slow magnetoacoustic waves as found by \citet{1998ApJ...501L.217D, 2006A&A...452.1059O, 2007A&A...463..713O}. 
Recently, \citet{2009A&A...493..251G} have reported on the detection
of these waves on the disk part of the coronal hole. They also find a different nature of compressional waves in bright (network) and dark (internetwork) regions of polar coronal hole (here after pCH). In this short contribution we extend the same analysis for another dataset. For further details please refer to \citet{2009A&A...493..251G}.

\section{Observations and data analysis}     \label{gupta-sec:observations}
Data used for this analysis was taken on 25th February 1997, from
00:00 UT to 13:59 UT with the 1$\times$300 \arcsec slit on SUMER, and
with an exposure time of 60 sec, in N~{\sc iv} 765 \AA\ and Ne~{\sc
  viii} 770 \AA\ lines in a southern coronal hole. For details on data
reduction please refer to \citet{2009A&A...493..251G}. 
 The  Chromosphere and Transition Region show a network (and cell/internetwork) pattern, with intensity enhancements in the network boundaries. Presumably, the magnetic field is predominantly concentrated on such network boundaries and, within coronal holes, the footpoints of coronal funnels
emanate from these network boundaries. As the duration of
the observation in this dataset is very long, the location of bright and dark pixels along the slit will change with time. For this reason, we have analysed the whole dataset pixel by pixel and timeframe by timeframe.
\begin{figure*}[hbp]
\centering
\hspace*{-0.4cm}\includegraphics[angle=90, width=7cm]{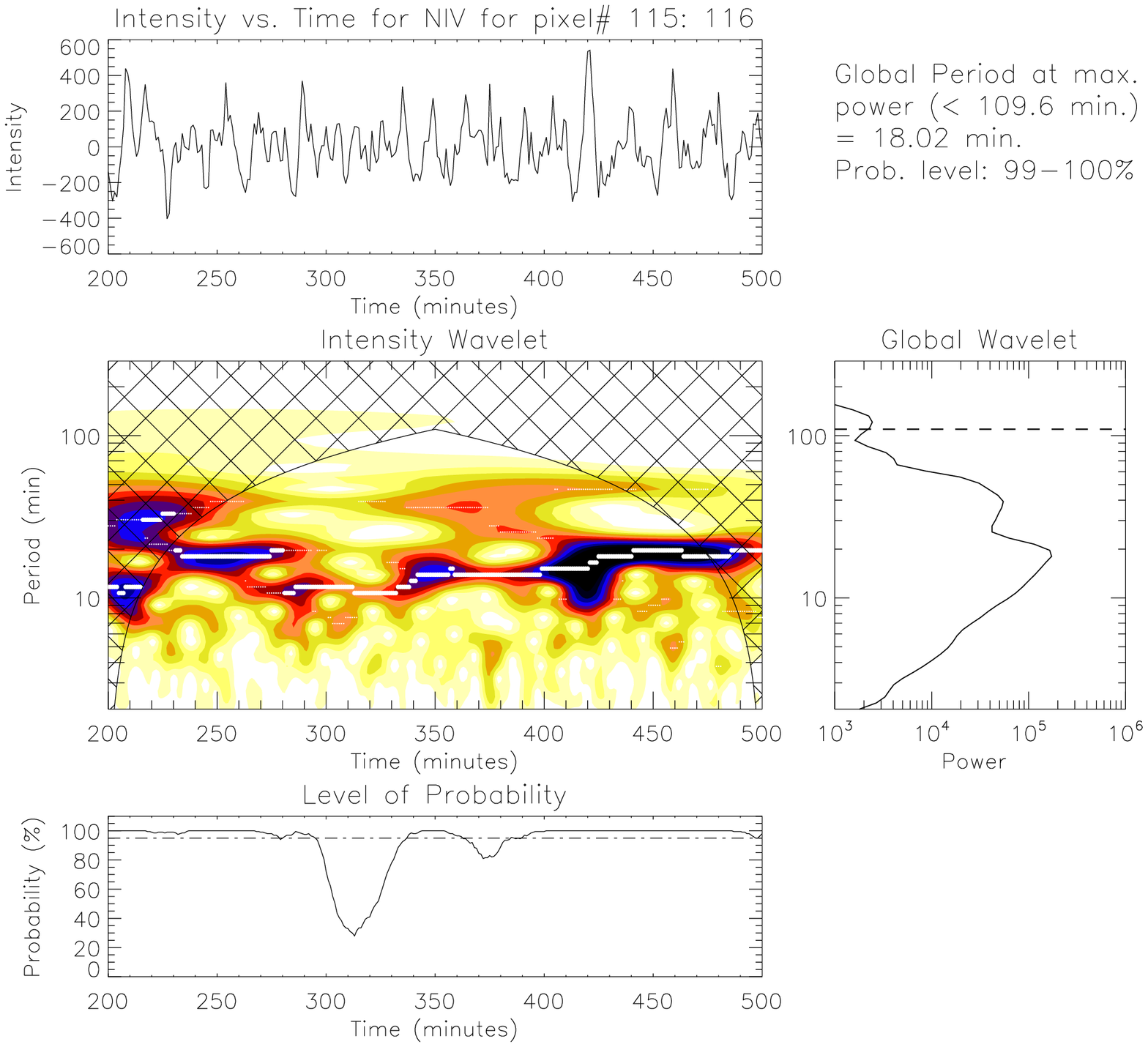}\hspace*{-1.0cm}{\includegraphics[angle=90, width=7cm]{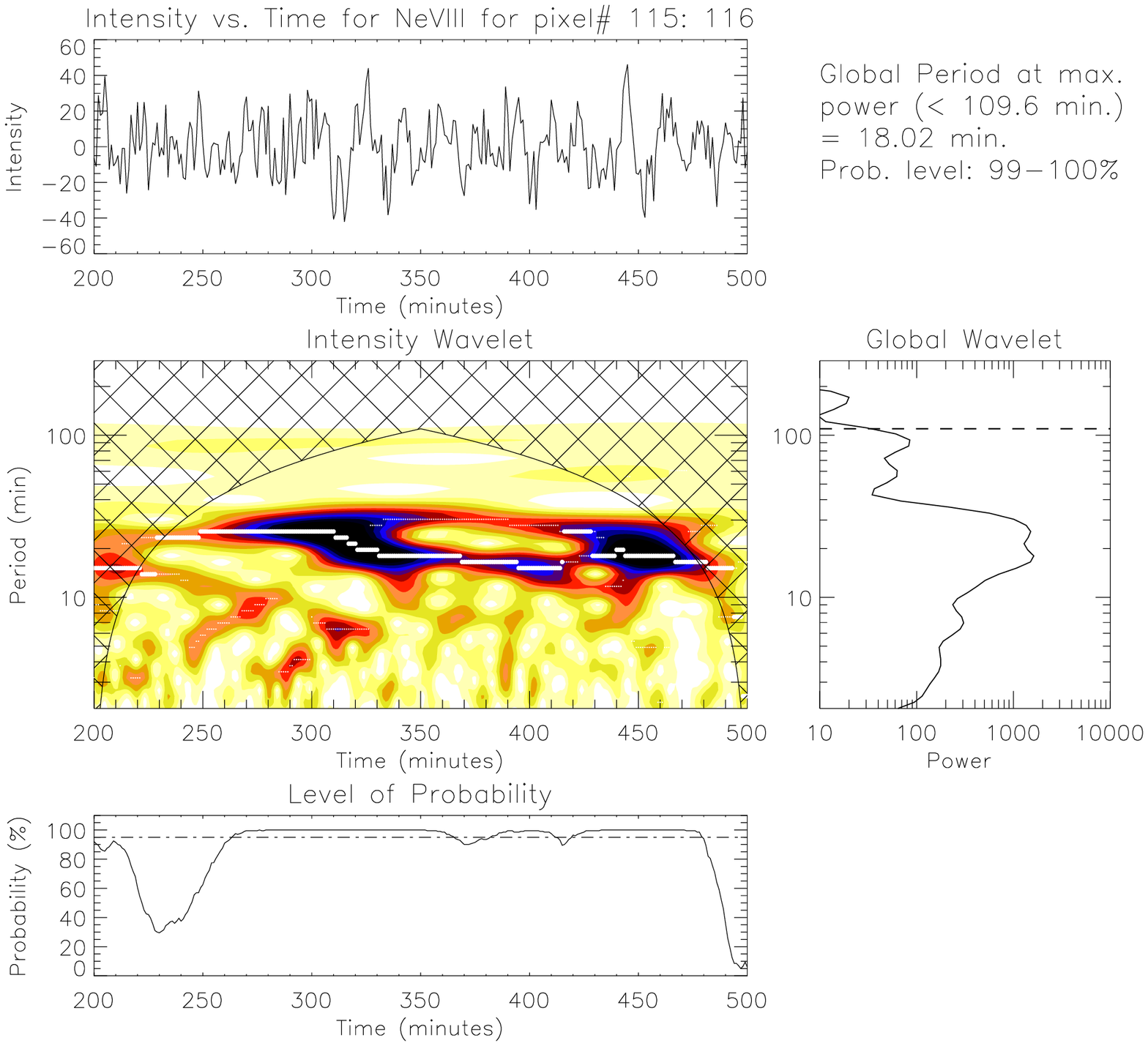}}
 \caption{The wavelet result for a bright location (pixel no. $=115-116$ which corresponds to Y$\approx-915\arcsec$) in the  N~{\sc iv} and Ne~{\sc viii} intensity. In each set the
 top panels show the relative (background trend removed) radiant
 flux, the central panels show the colour inverted wavelet power
 spectrum, the bottom panels show the variation of the probability
 estimate associated with the maximum power in the wavelet power
 spectrum (marked with white lines), and the right middle panels show the
 global (averaged over time) wavelet power spectrum. Above the global
 wavelet the period, measured from the maximum power from the global wavelet,
 together with probability estimate, is printed.}
 \label{giresh-fig:oscillation}
\end{figure*}
For example, for one time instance we first calculate the average intensity along the slit. All pixels having an intensity higher than 1.25 times this average intensity were chosen as bright pixels. If, this pixel is bright for at least 60 min (or 60 timeframes), then that pixel is considered to be a bright (`network') location over that time interval. The bright pixel identification  is done only for the
low temperature N~{\sc iv}  line; the network pixels obtained from it are assumed to be the same in the higher temperature Ne~{\sc viii} line.

\section{Results and Discussions}    \label{gupta-sec:results}
In Fig.~\ref{giresh-fig:oscillation}, we show a representative example
of the type of oscillations measured in a bright region of the pCH
using wavelets, which provides information on the temporal variation
of a signal \citep{1998BAMS...79...61T}.
Further details on the wavelet analysis used here can be found in
\cite{2009A&A...493..251G, 2001A&A...368.1095O} and reference
therein. From Fig.~\ref{giresh-fig:oscillation}, we can see oscillations
of $\approx$ 18 min period in both lines at the same location. This
gives an indication that these two layers are linked. A propagating
wave passing from one layer to the other could explain this finding.
 To test this hypothesis and to find the nature of the propagating waves, we measure the phase
delays in intensity and also in the LOS relative velocity
between both lines for each of the measurable pixels along the
slit for a full frequency range. This work has followed the method
used by \citet{1979ApJ...229.1147A, 2006A&A...452.1059O} and
\citet{2009A&A...493..251G} which make use of the basic equation of phase delay,
$ \Delta\phi = 2\pi\textit{f}T $
where $\textit{f}$ is the frequency and T the time delay in
seconds. From this relation, we expect the phase difference will vary linearly with \textit{f}, and will
change by $360\deg$ over frequency intervals of $\Delta \textit{f} =
1/T$. This will give rise to parallel lines in $\Delta\phi$
vs. $\textit{f}$ plots at fixed frequency intervals ($\Delta
\textit{f} =1/T$), corresponding to a fixed time delay T. 
\begin{figure*}[h]
 \centering
\hspace*{-0.4cm}\includegraphics[angle=90, width=6.3cm]{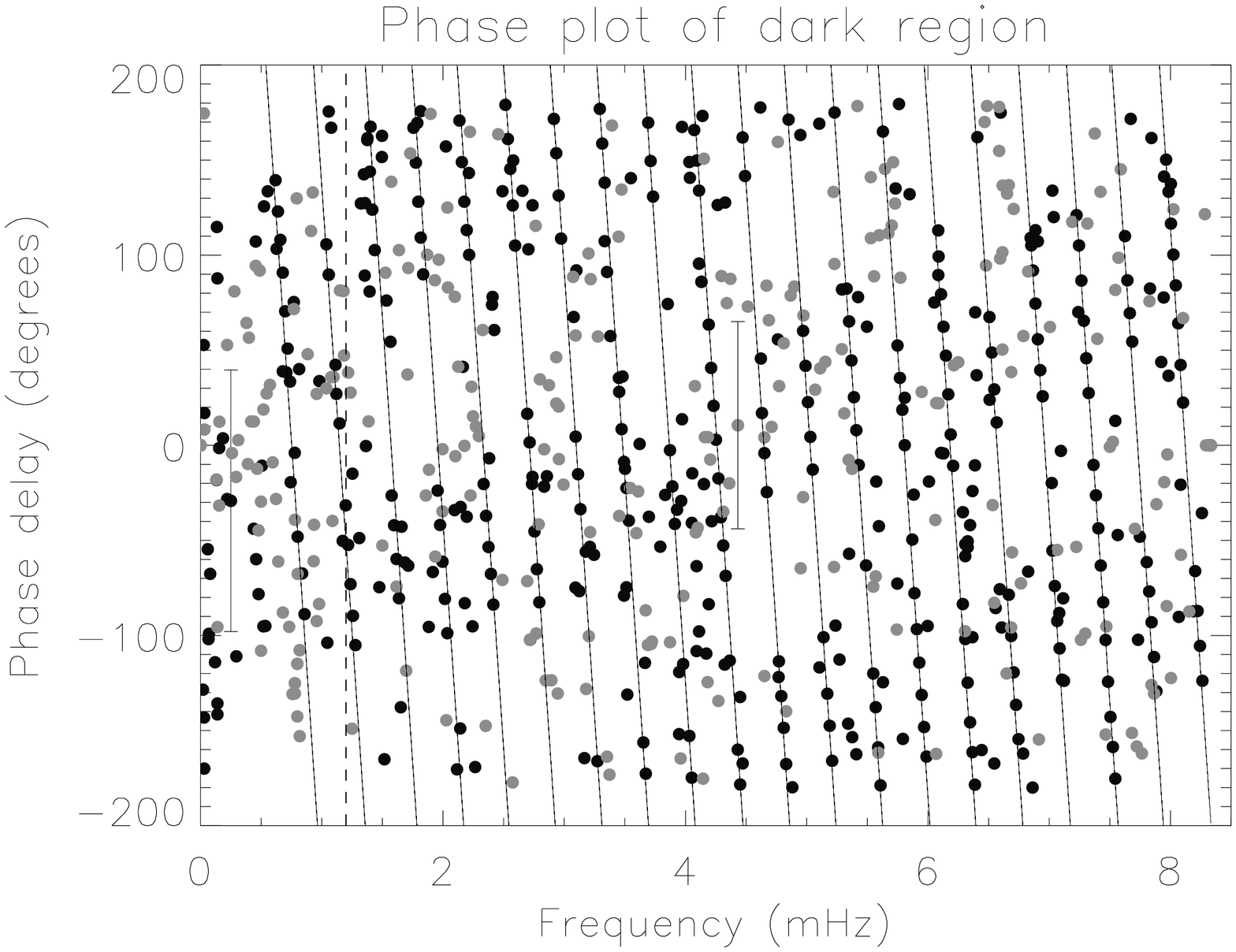}{\includegraphics[angle=90, width=6.3cm]{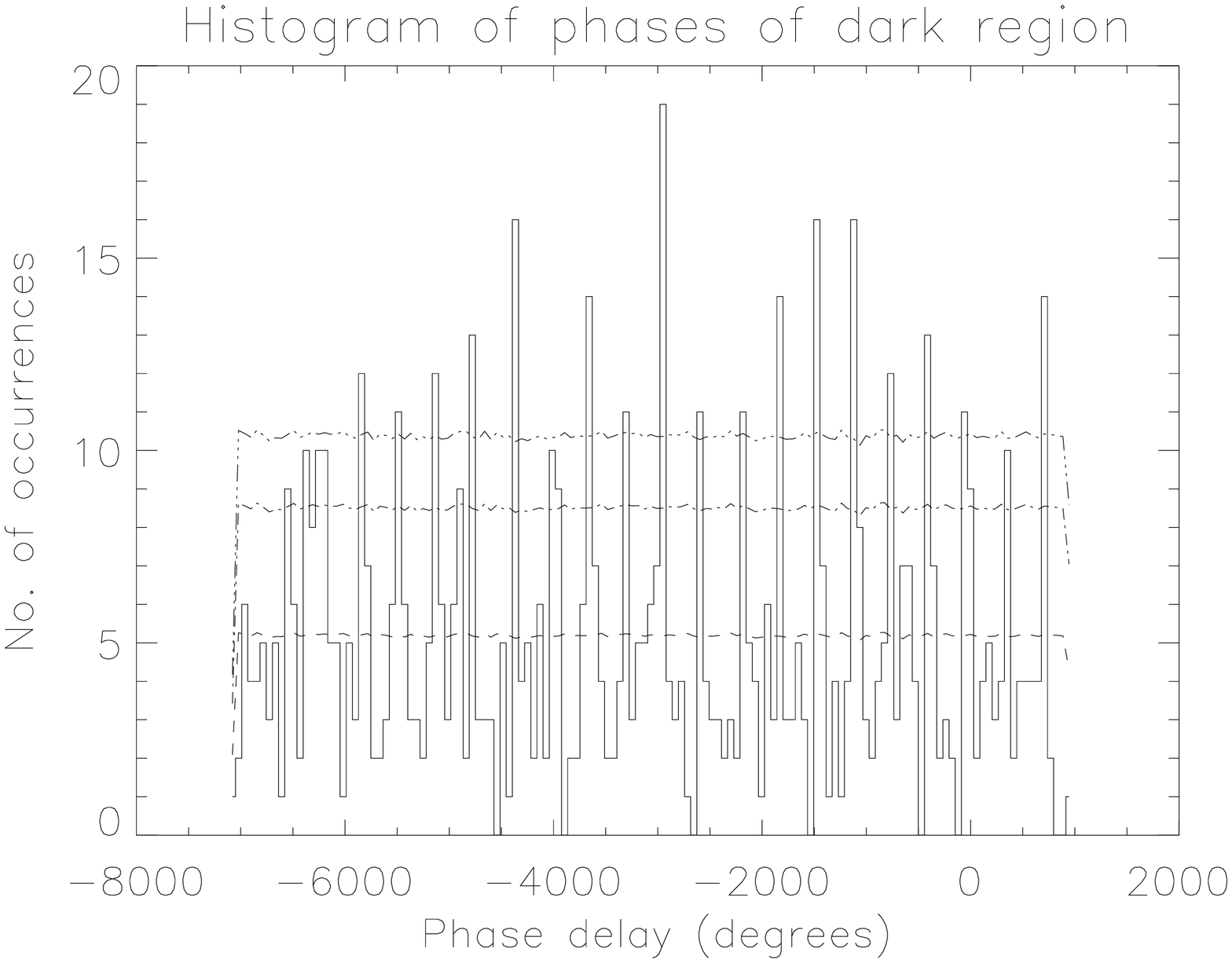}}
 \caption{Left panel: Phase delays measured between the oscillations in the
 spectroscopic line pair for dark locations . The phases
 from radiant flux oscillations are shown as the grey circle symbols
 while those from LOS velocities are shown as the black circle
 symbols. Overplotted on each figure are black  parallel lines,
 corresponding to fixed time delays. The vertical dashed line drawn at 1.2 mHz indicates that some phase values below this could be affected by solar rotation. Representative errors on the phase measurements are indicated by
 the error bars. Right panel: Histogram showing the distribution of phase delay
  measurements as a function of frequency for the dark locations. The horizontal dotted, dot-dashed and long dashed lines show the 68.3\%
  (i.e.,1$\sigma$), 90\% ($\approx$1.64$\sigma$) and
  95\% ($\approx$2$\sigma$) confidence levels, respectively,
  calculated using Monte-Carlo simulations with 5000 permutations.}
 \label{gupta-fig:phase_histo}
\end{figure*}
The method used here to identify the parallel lines in the phase plot
(left panel of Fig.~\ref{gupta-fig:phase_histo}) and the histogram
(right panel of Fig.~\ref{gupta-fig:phase_histo}), generated to
obtain the spacing between the lines and their significance level, is
discussed in greater detail in \citet{2009A&A...493..251G}.  In the
bright location, we didn't have enough statistics to come to a
reliable conclusion. Because of space constraints we do not show the
corresponding plots for the bright region here. Whereas oscillations were detected within the dark internetwork locations also and the statistics is much better and we show the results here in Fig.~\ref{gupta-fig:phase_histo}.

In the dark locations of pCH (left panel of Fig.~\ref{gupta-fig:phase_histo}), we measured a
time delay of $-2607 \pm 491.2$ s between the two lines, indicating upwardly propagating
waves. Notice the many closely spaced,
steeply sloped, parallel phase lines that correspond to this long time
delay. In right panel of Fig.~\ref{gupta-fig:phase_histo} these  parallel phase lines
correspond to significant peaks above 95\% significance in the
histogram. Here, the peaks are equally spaced at the phase
difference of $360\deg$ that would be expected for unimpeded
propagating waves.

The measured time delays from the pCH may be used to
estimate propagation speeds for the waves assumed to be causing the
oscillations. In order to calculate the propagation speeds, one needs
information on the height difference in the atmosphere between the
different lines in the line pair. The height difference of 4095 km has
been calculated using the limb brightening technique as described in
\citet{2006A&A...452.1059O, 2009A&A...493..251G}. Using this height
difference and the measured time delay, we find a speed of $-1.6 \pm
0.3$ km s$^{-1}$ for the dark region which is less than the acoustic
velocity at that height. Hence, the identified waves can be termed as
slow magento-acoustic waves. These waves are rather slow and may not
carry enough energy flux for the acceleration of the solar wind.

\bibliographystyle{rr-assp.bst}
\bibliography{gupta.bib}

\end{document}

%% file: rr-latexdefs.tex

\def\thisvolume{these proceedings}

\def\aj{{AJ}}			
\def\araa{{ARA\&A}}		
\def\apj{{ApJ}}			
\def\apjl{{ApJ}}		
\def\apjs{{ApJS}}		
\def\ao{{Appl.\ Optics}} 
\def\apss{{Ap\&SS}}		
\def\aap{{A\&A}}		
\def\aapr{{A\&A~Rev.}}		
\def\aaps{{A\&AS}}		
\def\an{{Astron.\ Nachrichten}}
\def\aspcs{{ASP Conf.\ Ser.}}
\def\azh{{AZh}}			
\def\baas{{BAAS}}		
\def\jrasc{{JRASC}}		
\def\memras{{MmRAS}}		
\def\mnras{{MNRAS}}
\def\nat{{Nat}}		
\def\pra{{Phys.\ Rev.\ A}} 
\def\prb{{Phys.\ Rev.\ B}}		
\def\prc{{Phys.\ Rev.\ C}}		
\def\prd{{Phys.\ Rev.\ D}}		
\def\prl{{Phys.\ Rev.\ Lett}}	
\def\pasp{{PASP}}
\def\pasj{{PASJ}}		
\def\qjras{{QJRAS}}
\def\science{{Sci}}		
\def\skytel{{S\&T}}		
\def\solphys{{Solar\ Phys.}} 
\def\sovast{{Soviet\ Ast.}}  
\def\ssr{{Space\ Sci.\ Rev.}}
\def\svassp{{Astrophys.\ Space Science Proc.}}
\def\zap{{ZAp}}			
\let\astap=\aap
\let\apjlett=\apjl
\let\apjsupp=\apjs

\def\ion#1#2{{\rm #1}\,{\uppercase{#2}}}  
\def\deg{\hbox{$^\circ$}}
\def\sun{\hbox{$\odot$}}
\def\earth{\hbox{$\oplus$}}
\def\la{\mathrel{\hbox{\rlap{\hbox{\lower4pt\hbox{$\sim$}}}\hbox{$<$}}}}
\def\ga{\mathrel{\hbox{\rlap{\hbox{\lower4pt\hbox{$\sim$}}}\hbox{$>$}}}}
\def\sq{\hbox{\rlap{$\sqcap$}$\sqcup$}}
\def\arcmin{\hbox{$^\prime$}}
\def\arcsec{\hbox{$^{\prime\prime}$}}
\def\fd{\hbox{$.\!\!^{\rm d}$}}
\def\fh{\hbox{$.\!\!^{\rm h}$}}
\def\fm{\hbox{$.\!\!^{\rm m}$}}
\def\fs{\hbox{$.\!\!^{\rm s}$}}
\def\fdg{\hbox{$.\!\!^\circ$}}
\def\farcm{\hbox{$.\mkern-4mu^\prime$}}
\def\farcs{\hbox{$.\!\!^{\prime\prime}$}}
\def\fp{\hbox{$.\!\!^{\scriptscriptstyle\rm p}$}}
\def\micron{\hbox{$\mu$m}}
\def\onehalf{\hbox{$\,^1\!/_2$}}	
\def\onethird{\hbox{$\,^1\!/_3$}}
\def\twothirds{\hbox{$\,^2\!/_3$}}
\def\onequarter{\hbox{$\,^1\!/_4$}}
\def\threequarters{\hbox{$\,^3\!/_4$}}
\def\ubv{\hbox{$U\!BV$}}		
\def\ubvr{\hbox{$U\!BV\!R$}}		
\def\ubvri{\hbox{$U\!BV\!RI$}}		
\def\ubvrij{\hbox{$U\!BV\!RI\!J$}}		
\def\ubvrijh{\hbox{$U\!BV\!RI\!J\!H$}}		
\def\ubvrijhk{\hbox{$U\!BV\!RI\!J\!H\!K$}}		
\def\ub{\hbox{$U\!-\!B$}}		
\def\bv{\hbox{$B\!-\!V$}}		
\def\vr{\hbox{$V\!-\!R$}}		
\def\ur{\hbox{$U\!-\!R$}}


\def\labelitemi{{\bf --}}  

\def\rmit#1{{\it #1}}              
\def\rmit#1{{\rm #1}}              
\def\etal{\rmit{et al.}}           
\def\etc{\rmit{etc.}}           
\def\ie{\rmit{i.e.,}}              
\def\eg{\rmit{e.g.,}}              
\def\cf{cf.}                       
\def\viz{\rmit{viz.}}
\def\vs{\rmit{vs.}}

\def\rot{\hbox{\rm rot}}
\def\div{\hbox{\rm div}}
\def\lesssim{\mathrel{\hbox{\rlap{\hbox{\lower4pt\hbox{$\sim$}}}\hbox{$<$}}}}
\def\gtrsim{\mathrel{\hbox{\rlap{\hbox{\lower4pt\hbox{$\sim$}}}\hbox{$>$}}}}
\def\dif{\: {\rm d}}                       
\def\ep{\:{\rm e}^}                        
\def\dash{\hbox{$\,-\,$}}                  
\def\is{\!=\!}                             

\def\starname#1#2{${#1}$\,{\rm {#2}}}  
\def\Teff{\hbox{$T_{\rm eff}$}}   

\def\kms{\hbox{km$\;$s$^{-1}$}}
\def\Mxcm{\hbox{Mx\,cm$^{-2}$}}    

\def\Bapp{\hbox{$B_{\rm app}$}}    

\def\komega{($k, \omega$)}                 
\def\kf{($k_h,f$)}                         
\def\VminI{\hbox{$V\!\!-\!\!I$}}           
\def\IminI{\hbox{$I\!\!-\!\!I$}}           
\def\VminV{\hbox{$V\!\!-\!\!V$}}           
\def\Xt{\hbox{$X\!\!-\!t$}}                

\def\level #1 #2#3#4{$#1 \: ^{#2} \mbox{#3} ^{#4}$}   

\def\specchar#1{\uppercase{#1}}    
\def\AlI{\mbox{Al\,\specchar{i}}}  
\def\BI{\mbox{B\,\specchar{i}}} 
\def\BII{\mbox{B\,\specchar{ii}}}  
\def\BaI{\mbox{Ba\,\specchar{i}}}  
\def\BaII{\mbox{Ba\,\specchar{ii}}} 
\def\CI{\mbox{C\,\specchar{i}}} 
\def\CII{\mbox{C\,\specchar{ii}}} 
\def\CIII{\mbox{C\,\specchar{iii}}} 
\def\CIV{\mbox{C\,\specchar{iv}}} 
\def\CaI{\mbox{Ca\,\specchar{i}}} 
\def\CaII{\mbox{Ca\,\specchar{ii}}} 
\def\CaIII{\mbox{Ca\,\specchar{iii}}} 
\def\CoI{\mbox{Co\,\specchar{i}}} 
\def\CrI{\mbox{Cr\,\specchar{i}}} 
\def\CriI{\mbox{Cr\,\specchar{ii}}} 
\def\CsI{\mbox{Cs\,\specchar{i}}} 
\def\CsII{\mbox{Cs\,\specchar{ii}}} 
\def\CuI{\mbox{Cu\,\specchar{i}}} 
\def\FeI{\mbox{Fe\,\specchar{i}}} 
\def\FeII{\mbox{Fe\,\specchar{ii}}} 
\def\FeIX{\mbox{Fe\,\specchar{ix}}}
\def\FeX{\mbox{Fe\,\specchar{x}}}
\def\FeXVI{\mbox{Fe\,\specchar{xvi}}}
\def\FrI{\mbox{Fr\,\specchar{i}}}
\def\HI{\mbox{H\,\specchar{i}}} 
\def\HII{\mbox{H\,\specchar{ii}}} 
\def\Hmin{\hbox{\rmH$^{^{_{\scriptstyle -}}}$}}      
\def\Hemin{\hbox{{\rm He}$^{^{_{\scriptstyle -}}}$}} 
\def\HeI{\mbox{He\,\specchar{i}}} 
\def\HeII{\mbox{He\,\specchar{ii}}} 
\def\HeIII{\mbox{He\,\specchar{iii}}} 
\def\KI{\mbox{K\,\specchar{i}}} 
\def\KII{\mbox{K\,\specchar{ii}}} 
\def\KIII{\mbox{K\,\specchar{iii}}} 
\def\LiI{\mbox{Li\,\specchar{i}}} 
\def\LiII{\mbox{Li\,\specchar{ii}}} 
\def\LiIII{\mbox{Li\,\specchar{iii}}} 
\def\MgI{\mbox{Mg\,\specchar{i}}} 
\def\MgII{\mbox{Mg\,\specchar{ii}}} 
\def\MgIII{\mbox{Mg\,\specchar{iii}}} 
\def\MnI{\mbox{Mn\,\specchar{i}}} 
\def\NI{\mbox{N\,\specchar{i}}}
\def\NaI{\mbox{Na\,\specchar{i}}}
\def\NaII{\mbox{Na\,\specchar{ii}}}
\def\NaIII{\mbox{Na\,\specchar{iii}}} 
\def\NiI{\mbox{Ni\,\specchar{i}}} 
\def\NiII{\mbox{Ni\,\specchar{ii}}}
\def\NiIII{\mbox{Ni\,\specchar{iii}}} 
\def\OI{\mbox{O\,\specchar{i}}} 
\def\OVI{\mbox{O\,\specchar{vi}}}
\def\RbI{\mbox{Rb\,\specchar{i}}} 
\def\SII{\mbox{S\,\specchar{ii}}} 
\def\SiI{\mbox{Si\,\specchar{i}}} 
\def\SiII{\mbox{Si\,\specchar{ii}}} 
\def\SrI{\mbox{Sr\,\specchar{i}}}
\def\SrII{\mbox{Sr\,\specchar{ii}}}
\def\TiI{\mbox{Ti\,\specchar{i}}} 
\def\TiII{\mbox{Ti\,\specchar{ii}}} 
\def\TiIII{\mbox{Ti\,\specchar{iii}}} 
\def\TiIV{\mbox{Ti\,\specchar{iv}}} 
\def\VI{\mbox{V\,\specchar{i}}} 
\def\HtwoO{\mbox{H$_2$O}}        
\def\Otwo{\mbox{O$_2$}}          

\def\Halpha{\mbox{H\hspace{0.1ex}$\alpha$}} 
\def\Ha{\mbox{H\hspace{0.2ex}$\alpha$}}
\def\Hbeta{\mbox{H\hspace{0.2ex}$\beta$}}
\def\Hgamma{\mbox{H\hspace{0.2ex}$\gamma$}}
\def\Hdelta{\mbox{H\hspace{0.2ex}$\delta$}}
\def\Hepsilon{\mbox{H\hspace{0.2ex}$\epsilon$}}
\def\Hzeta{\mbox{H\hspace{0.2ex}$\zeta$}}
\def\Lyalpha{\mbox{Ly$\hspace{0.2ex}\alpha$}}
\def\Lybeta{\mbox{Ly$\hspace{0.2ex}\beta$}}
\def\Lygamma{\mbox{Ly$\hspace{0.2ex}\gamma$}}
\def\Lycont{\mbox{Ly\hspace{0.2ex}{\small cont}}}
\def\Baalpha{\mbox{Ba$\hspace{0.2ex}\alpha$}}
\def\Babeta{\mbox{Ba$\hspace{0.2ex}\beta$}}
\def\Bacont{\mbox{Ba\hspace{0.2ex}{\small cont}}}
\def\Paalpha{\mbox{Pa$\hspace{0.2ex}\alpha$}}
\def\Bralpha{\mbox{Br$\hspace{0.2ex}\alpha$}}

\def\NaD{\mbox{Na\,\specchar{i}\,D}}    
\def\NaDone{\mbox{Na\,\specchar{i}\,\,D$_1$}}
\def\NaDtwo{\mbox{Na\,\specchar{i}\,\,D$_2$}}
\def\NaID{\mbox{Na\,\specchar{i}\,\,D}}
\def\NaIDone{\mbox{Na\,\specchar{i}\,\,D$_1$}}
\def\NaIDtwo{\mbox{Na\,\specchar{i}\,\,D$_2$}}
\def\Done{\mbox{D$_1$}}
\def\Dtwo{\mbox{D$_2$}}

\def\Mgbone{\mbox{Mg\,\specchar{i}\,b$_1$}}
\def\Mgbtwo{\mbox{Mg\,\specchar{i}\,b$_2$}}
\def\Mgbthree{\mbox{Mg\,\specchar{i}\,b$_3$}}
\def\MgIb{\mbox{Mg\,\specchar{i}\,b}}
\def\MgIbone{\mbox{Mg\,\specchar{i}\,b$_1$}}
\def\MgIbtwo{\mbox{Mg\,\specchar{i}\,b$_2$}}
\def\MgIbthree{\mbox{Mg\,\specchar{i}\,b$_3$}}

\def\CaIIK{\mbox{Ca\,\specchar{ii}\,K}}       
\def\CaIIH{\mbox{Ca\,\specchar{ii}\,H}}
\def\CaIIHK{\mbox{Ca\,\specchar{ii}\,H\,\&\,K}}
\def\HK{\mbox{H\,\&\,K}}
\def\Kthree{\mbox{K$_3$}}      
\def\Hthree{\mbox{H$_3$}}
\def\Ktwo{\mbox{K$_2$}}
\def\Htwo{\mbox{H$_2$}}
\def\Kone{\mbox{K$_1$}}     
\def\Hone{\mbox{H$_1$}}     
\def\KtwoV{\mbox{K$_{2V}$}}
\def\KtwoR{\mbox{K$_{2R}$}}
\def\KoneV{\mbox{K$_{1V}$}}
\def\KoneR{\mbox{K$_{1R}$}}
\def\HtwoV{\mbox{H$_{2V}$}}
\def\HtwoR{\mbox{H$_{2R}$}}
\def\HoneV{\mbox{H$_{1V}$}}
\def\HoneR{\mbox{H$_{1R}$}}

\def\hk{\mbox{h\,\&\,k}}
\def\kthree{\mbox{k$_3$}}    
\def\hthree{\mbox{h$_3$}}
\def\ktwo{\mbox{k$_2$}}
\def\htwo{\mbox{h$_2$}}
\def\kone{\mbox{k$_1$}}     
\def\hone{\mbox{h$_1$}}     
\def\ktwoV{\mbox{k$_{2V}$}}
\def\ktwoR{\mbox{k$_{2R}$}}
\def\koneV{\mbox{k$_{1V}$}}
\def\koneR{\mbox{k$_{1R}$}}
\def\htwoV{\mbox{h$_{2V}$}}
\def\htwoR{\mbox{h$_{2R}$}}
\def\honeV{\mbox{h$_{1V}$}}
\def\honeR{\mbox{h$_{1R}$}}

%% file: gupta.bbl
\begin{thebibliography}{11}
\expandafter\ifx\csname natexlab\endcsname\relax\def\natexlab#1{#1}\fi

\bibitem[{{Athay} \& {White}(1979)}]{1979ApJ...229.1147A}
{Athay}, R.~G. {White}, O.~R. 1979, \apj, 229, 1147

\bibitem[{{Banerjee} {et~al.}(2001){Banerjee}, {O'Shea}, {Doyle}, \&
  {Goossens}}]{2001A&A...380L..39B}
{Banerjee}, D., {O'Shea}, E., {Doyle}, J.~G., {Goossens}, M. 2001, \aap, 380,
  L39

\bibitem[{{Deforest} \& {Gurman}(1998)}]{1998ApJ...501L.217D}
{Deforest}, C.~E. {Gurman}, J.~B. 1998, \apjl, 501, L217+

\bibitem[{{Gupta} {et~al.}(2009){Gupta}, {O'Shea}, {Banerjee}, {Popescu}, \&
  {Doyle}}]{2009A&A...493..251G}
{Gupta}, G.~R., {O'Shea}, E., {Banerjee}, D., {Popescu}, M., {Doyle}, J.~G.
  2009, \aap, 493, 251

\bibitem[{{Ofman} {et~al.}(1997){Ofman}, {Romoli}, {Poletto}, {Noci}, \&
  {Kohl}}]{1997ApJ...491L.111O}
{Ofman}, L., {Romoli}, M., {Poletto}, G., {Noci}, G., {Kohl}, J.~L. 1997,
  \apjl, 491, L111+

\bibitem[{{Ofman} {et~al.}(2000){Ofman}, {Romoli}, {Poletto}, {Noci}, \&
  {Kohl}}]{2000ApJ...529..592O}
{Ofman}, L., {Romoli}, M., {Poletto}, G., {Noci}, G., {Kohl}, J.~L. 2000, \apj,
  529, 592

\bibitem[{{O'Shea} {et~al.}(2006){O'Shea}, {Banerjee}, \&
  {Doyle}}]{2006A&A...452.1059O}
{O'Shea}, E., {Banerjee}, D., {Doyle}, J.~G. 2006, \aap, 452, 1059

\bibitem[{{O'Shea} {et~al.}(2007){O'Shea}, {Banerjee}, \&
  {Doyle}}]{2007A&A...463..713O}
{O'Shea}, E., {Banerjee}, D., {Doyle}, J.~G. 2007, \aap, 463, 713

\bibitem[{{O'Shea} {et~al.}(2001){O'Shea}, {Banerjee}, {Doyle}, {Fleck}, \&
  {Murtagh}}]{2001A&A...368.1095O}
{O'Shea}, E., {Banerjee}, D., {Doyle}, J.~G., {Fleck}, B., {Murtagh}, F. 2001,
  \aap, 368, 1095

\bibitem[{{Popescu} {et~al.}(2005){Popescu}, {Banerjee}, {O'Shea}, {Doyle}, \&
  {Xia}}]{2005A&A...442.1087P}
{Popescu}, M.~D., {Banerjee}, D., {O'Shea}, E., {Doyle}, J.~G., {Xia}, L.~D.
  2005, \aap, 442, 1087

\bibitem[{{Torrence} \& {Compo}(1998)}]{1998BAMS...79...61T}
{Torrence}, C. {Compo}, G.~P. 1998, Bulletin of the American Meteorological
  Society, 79, 61

\end{thebibliography}
